\documentclass[%
 reprint,
superscriptaddress,
 amsmath,amssymb,
 aps,
prb,
longbibliography
]{revtex4-2}

\usepackage{graphicx}% Include figure files
\usepackage{dcolumn}% Align table columns on decimal point
\usepackage{bm}% bold math
\usepackage{booktabs}
\usepackage{siunitx}
\usepackage{float}
\usepackage{natbib}
\usepackage{makecell}
\usepackage{balance}
\usepackage{color, amsmath}
\usepackage{notoccite}

\usepackage{hyperref}% add hypertext capabilities
\hypersetup{colorlinks=false, pdfborder = {0 0 0}}
\begin{document}
%\renewcommand{\BHBI}{-}

%\preprint{APS/123-QED}

\title{Strong-coupling corrections to hard domain walls in superfluid $^3$He-B}

\author{M. J. Rudd}
\affiliation{Department of Physics, University of Alberta, Edmonton, Alberta, T6G 2E1, Canada}

\author{P. Senarath Yapa}
\affiliation{Department of Physics, University of Alberta, Edmonton, Alberta, T6G 2E1, Canada}

\author{A. J. Shook}
\affiliation{Department of Physics, University of Alberta, Edmonton, Alberta, T6G 2E1, Canada}

\author{J. Maciejko}
\email{maciejko@ualberta.ca}
\affiliation{Department of Physics, University of Alberta, Edmonton, Alberta, T6G 2E1, Canada}
\affiliation{Theoretical Physics Institute, University of Alberta, Edmonton, Alberta, T6G 2E1, Canada}

\author{J. P. Davis}
\email{jdavis@ualberta.ca}
\affiliation{Department of Physics, University of Alberta, Edmonton, Alberta, T6G 2E1, Canada}

\date{\today}

\begin{abstract}
Domain walls in superfluid $^3$He-$B$ have gained renewed interest in light of experimental progress on confining helium in nanofabricated geometries.  Here, we study the effect of strong-coupling corrections on domain wall width and interfacial tension by determining self-consistent solutions to spatially-dependent Ginzburg-Landau equations. We find that the formation of domain walls is generally energetically favored in strong coupling over weak coupling. Calculations were performed over a wide range of temperatures and pressures, showing decreasing interface energy with increasing temperature and pressure. This has implications for the observability of such domain walls in $^3$He-$B$, which are of both fundamental interest and form the basis for spatially-modulated pair-density wave states, when stabilized by strong confinement.
\end{abstract}

\maketitle

\section{Introduction}
Superfluid $^3$He, due to its $p$-wave, spin triplet wavefunction, possesses a rich landscape of spontaneously broken symmetries. In addition to the $\textsf{U}(1)_\textrm{N}$ gauge symmetry breaking expected for off-diagonal long-range order in superfluids, broken spin and orbital rotation symmetries $\textsf{SO}(3)_\textrm{S}$ $\times$ $\textsf{SO}(3)_\textrm{L}$ suggest numerous configurations of long-range order in the superfluid system. In turn, this implies the existence of many possible superfluid phases of $^3$He. Despite this, in isotropic, homogeneous (bulk) superfluid $^3$He only two are observed: the $A$ and $B$ phases \cite{osheroff1972-1, osheroff1972-2, anderson1961, balian1963, anderson1973, leggett1975, wheatley1975}. By introducing anisotropies not seen in the bulk, it is possible to access additional stable states of superfluid $^3$He \cite{pollanen2012, li2013, dmitriev2015, dmitriev2020}. Of particular interest are the effects of spatial confinement of the fluid in geometries of scale $D$ on the order of its temperature-dependent pair coherence length $\xi(T)$ \cite{vorontsov2007, wiman2015, wiman2016}.

As we will see below,  in a slab confined in $z$, scattering of Cooper pairs at a surface parallel to the $xy$-plane suppresses the $z$-component of the order parameter, leading to the appearance of a planar-distorted $B$-phase in the phase diagram \cite{vorontsov2018}. This surface pair-breaking comes with an energy cost incurred along the surface. This may be partially offset via a periodic arrangement of $^3$He-$B$ domains, whose formation was predicted by Vorontsov and Sauls in Ref.~\cite{vorontsov2007}. Originally conceived of as a stripe phase, one can generally consider a pair density wave state (PDW) \cite{shook2020}, where domains with alternating order-parameter components are arranged such that the pair-breaking cost is eliminated near the domain walls in exchange for the (lower) energy cost of the wall between the degenerate domains.

The prediction of a spatially-modulated phase sparked significant experimental investigation of $^3$He under confinement, using varied techniques including microfabricated fourth sound resonators \cite{shook2020}, nuclear magnetic resonance \cite{levitin2013, levitin2019}, shear micromechanical resonators \cite{zheng2016}, and torsional oscillators \cite{zhelev2017}. Significantly, two sequential first-order phase transitions have been observed \cite{shook2020}, and NMR studies have observed simultaneous evidence of two different planar-distorted $B$ phases \cite{levitin2019}. These results are indicative of the existence of a PDW state, but are yet unable to determine its spatial arrangement, with the authors of Ref.~\cite{levitin2019} proposing an alternative of textural domain walls, also called soft domain walls, and recent theoretical results concluding a triangular PDW is preferred \cite{yapa2021}. 

Importantly, the stripe phase was originally predicted in a study of hard domain walls in $^3$He-$B$ \cite{vorontsov2005} using the weak-coupling limit of Bardeen-Cooper-Schrieffer (BCS) theory.  This has led some to suggest that strong-coupling corrections could decrease the stability of the stripe phase \cite{levitin2019}. In response, strong-coupling calculations have been performed, which do show an increased stability of the $A$-phase and a suppression of the stripe phase, yet still predict stripe phase formation at low pressure and achievable temperatures \cite{wiman2016}. If an alternate PDW state with domain walls is realized, such as the triangular PDW \cite{yapa2021}, there remains the question as to what effect strong-coupling corrections have on these domain walls.  Furthermore, the observation of these domain walls is of fundamental interest as the two-dimensional analog of a quantum vortex \cite{makinen2019}, with parallels in cosmology \cite{salomaa1988, winkelmann2006, volovik2003, volovik2020a, volovik2020b}.

In this work, we use numerical methods to determine self-consistent solutions of the superfluid order parameter near interfaces of degenerate regions of bulk superfluid $^3$He-$B$. These $B$-$B$ hard domain walls have been previously treated by Salomaa and Volovik \cite{salomaa1988}, as well as Vorontsov and Sauls \cite{vorontsov2005}, and numerically by Silveri et al.~\cite{silveri2014}, with the latter using a perturbation method to determine that there is only one possible $B$-$B$ interface stable in the bulk. Here, we closely follow the work of Ref.~\cite{silveri2014}, but now with the addition of experimental strong-coupling corrections to the Ginzburg-Landau parameters, and obtain self-consistent solutions for the spatially-modified order parameter. For each of the domain configurations computed in Ref.~\cite{silveri2014}, we examine the width of the domain wall, as well as the energy of domain wall formation. We determine that strong-coupling corrections result in an enlargement of the width of the domain wall structure along with a reduction in its energy. These changes are consistent across the phase diagram with the exception of a small region near saturated vapor pressure.

The rest of the manuscript proceeds as follows.  In Section II, we begin by reviewing the origin and classification of domain walls in $^3$He-$B$.  This is followed in Section III by a review of Ginzburg-Landau theory and the incorporation of strong-coupling corrections.  We also discuss the definition of the interfacial tension, and the stability of the domain wall solutions. In Section IV, we discuss the calculation results for the domain wall widths, and in Section V our results for the interfacial tension. In Section VI, we discuss how the one-dimensional solutions found here are applicable to domain walls in a PDW and finally, in Section VII, we conclude that the physically relevant strong-coupling corrections do not inhibit the formation of domain walls in $^3$He-$B$ and hence the formation of PDW states.

\section{Domain walls in bulk $^3$He}
Weak-coupling BCS theory extended to p-wave, spin-triplet pairing posits that the state which is isotropic in \textbf{k}-space will be energetically favored \cite{balian1963}, indicating that the $B$ phase, with its uniform energy gap, should be the dominant superfluid phase. Experimental data reveals that near $T_c$ and at high pressures, a region of $A$ phase forms, implying the necessity of corrections to weak-coupling at higher pressures. The $\mathbf{k}$-space isotropy of the $B$ phase gives an order parameter of the form:
\begin{align}
    A_{\mu i} &= \Delta e^{i\phi}R_{\mu i}(\mathbf{\hat{n}}, \theta), \label{eq:B-phase OP-rln}
\end{align}
where $\Delta$ is a real, positive amplitude, $\phi$ a real-valued phase, and $R_{\mu i}$ a proper rotation matrix \cite{viljas2002, silveri2014}. This state is degenerate with respect to multiplication by a constant phase factor, and by rotating spin relative to position space \cite{viljas2002, silveri2014}. Thus we may conceive of many possible degenerate $B$ phases in isolation.

We consider the characteristics of a domain wall between two degenerate, semi-infinite bulk volumes of superfluid $^3$He-$B$. We expect there to be a transition zone between these domains. Given appropriate boundary conditions, if a self-consistent solution of finite extent can be found, we define this as a domain wall. The width, $l$, of this domain wall is defined as the solution domain with nonzero gradient \footnote{Numerically, the gradient does not reach zero until the boundary condition. As such, we set a lower limit of $10^{-4}$, below which the gradient is taken to be zero for the purposes of defining $l$} and is on the order of $\xi_\Delta(T)$. We contrast this with a \textit{texture}, where we fail to obtain a converged solution even when arbitrarily expanding the size of the calculation domain.

Following the method of Ref.~\cite{viljas2002}, we use symmetry to determine and name the realizable $B$-$B$ interfaces. All external factors being equal, any observable characteristic of an interface should result from the configuration of the domains, which may be described in terms of invariant parameters. For our system symmetries, these invariants are:
\begin{align}\label{eq:phi-defns}
    \phi &= \phi^R - \phi^L, \quad \psi_\perp = R_{\mu z}^L R_{\mu z}^R,\\
    \psi_\parallel &= R^L_{\mu x}R^R_{\mu x} + R_{\mu y}^L R_{\mu y}^R,\nonumber
\end{align}
where we have used the superscripts $L, R$ to indicate order parameter values on the left and right sides of the interface, respectively. $\phi$ reflects the invariance under global phase shifts, and $\psi_\perp, \, \psi_\parallel$ arise from the invariance under global spin rotation combined with rotational symmetry about the interface normal \cite{viljas2002}. These invariants can be condensed further into dependence on two complex numbers:
\begin{align} \label{eq: a-b defns}
    a = e^{i\phi}\psi_\perp, \quad b= e^{i\phi}\psi_\parallel,
\end{align}
which fully define the $B$-$B$ domain wall configuration.
\begin{table}[b]
\caption{\label{table-order-parameters}Summary of domain wall configurations. The second column gives the boundary conditions imposed on the scaled order parameter $A_{\mu i}/\Delta_B$ on the right-hand side of the interface. In all cases, the left-hand side is represented by the identity matrix. All numerical values are for weak-coupling calculations.}
\begin{ruledtabular}
\begin{tabular}{ccccc}
\addlinespace[1ex]
Name &  $A_{\mu i}^R / \Delta_B$ & $l$ $(\xi_{\Delta})$ &  $\sigma \, (f_c^B \xi_{\Delta})$  & $\sigma_G / \sigma$\\
\addlinespace[1ex]
\colrule
\addlinespace[2ex]
BB12 & $\begin{bmatrix}
+1 & 0 & 0\\
0 & +1 & 0 \\
0 & 0 & +1
\end{bmatrix}$ & N/A & 0 & N/A\\
\addlinespace[1.5ex]
BB10 & $\begin{bmatrix}
-1 & 0 & 0\\
0 & +1 & 0 \\
0 & 0 & +1
\end{bmatrix}$ & 25.16 & 0.9006 & 0.4999 \\
\addlinespace[1.5ex]
BB$\overline{1}2$ & $\begin{bmatrix}
+1 & 0 & 0\\
0 & +1 & 0 \\
0 & 0 & -1
\end{bmatrix}$ & 32.90 & 1.5547 & 0.5000\\
\addlinespace[1.5ex]
BB$1\overline{2}$ & $\begin{bmatrix}
-1 & 0 & 0\\
0 & -1 & 0 \\
0 & 0 & +1
\end{bmatrix}$ & 27.80 & 2.1018 & 0.5001\\
\addlinespace[1.5ex]
BB$\overline{1}0$ & $\begin{bmatrix}
-1 & 0 & 0\\
0 & +1 & 0 \\
0 & 0 & -1
\end{bmatrix}$ & 32.54 & 2.7853 & 0.5000\\
\addlinespace[1.5ex]
BB$\overline{12}$ & $\begin{bmatrix}
-1 & 0 & 0\\
0 & -1 & 0 \\
0 & 0 & -1
\end{bmatrix}$ & 32.12 & 4.5754 & 0.5000\\
\end{tabular}
\end{ruledtabular}
\end{table}
We are concerned only with domain walls that are stable under static conditions, therefore we eliminate any values of $a$ and $b$ that result in spin or mass currents across the interface, although it would be interesting in future work to determine whether domain walls exist that are stabilized by flow in a dynamic system \cite{wu2013, shook2020}. From the restriction on zero mass current, we see that $a$ and $b$ must be real, with the spin current also vanishing in the cases $a=\pm 1$, $b=0$, $b=\pm 2$. By cycling over these $a$ and $b$ values we determine which interfaces may be realizable in the bulk, in the absence of external fields or currents \cite{silveri2014}. We performed calculations for each of the possible six domain wall configurations, which are listed in Table \ref{table-order-parameters}. For consistency, we continue to use the naming system proposed in Ref.~\cite{silveri2014} in which the interfaces are named as $BBab$, with negative values denoted by a bar over the number. We note that these configurations are equivalent to all possible permutations of the $B$-phase order parameter differing by a change of sign in one or more components, due to the equivalence between the $A_{xx}$ and $A_{yy}$ components of the order parameter demonstrated in (\ref{eq:phi-defns}) and (\ref{eq: a-b defns}).

An important consideration for each of these domain wall structures and their energetics is which components of the order parameter change sign. In this work, we use the nomenclature employed by Vorontsov \cite{vorontsov2018} and denote the structures as \textit{type-$x$} or \textit{type-$z$}. Mixed combinations are theoretically possible for more complex domain walls. The type-$z$ domain wall structure is defined by $\nabla_i A_{\mu i} \ne 0$. For our solutions with variation in $\hat{z}$, this means that $A_{\mu z}$ components change sign (hence type-$z$), with the type-$z$ interface the $BB\overline{1}2$ domain wall. The type-$x$ domain wall is defined by $\nabla_i A_{\mu j} \ne 0, \, i\ne j$, with the simplest such interface the $BB10$ domain wall. As will be shown in Section \ref{sec:GL-theory}, a rotation of the axes for 1D solutions is possible, provided the indices of $A_{\mu i}$ are changed accordingly. Therefore, although Table \ref{table-order-parameters} shows the $BB10$ domain wall having a sign change in $A_{xx}$ when $\hat{z}$ is the solution direction, an equivalent arrangement can have order parameter component $A_{z z}$ change sign so long as the solution is changing in $\hat{x}$ or $\hat{y}$ $(\nabla_{x,y} A_{\mu i}\ne 0)$. As mentioned above, the calculations described in Section \ref{sec:GL-theory} were performed for all domain wall configurations listed in Table \ref{table-order-parameters}. For conciseness, we have elected to present primarily the results of calculations on the $BB10$ and $BB\overline{1}2$ domain walls, as these are the simplest examples of the type-$x$ and type-$z$ domain walls, respectively. However, we would like to emphasize that the conclusions drawn from these results are consistent for all domain walls examined.

\section{Ginzburg-Landau theory and calculations}\label{sec:GL-theory}
We find self-consistent solutions for the order parameter using Ginzburg-Landau theory, modifying the weak coupling model for use at $P \in [0,34]$ bar and temperatures below the critical temperature $T_c$. The free energy functional of $^3$He, as a function of the order parameter $A_{\mu i}$, is $F=F_b + F_g$ with $F_b$ the bulk energy:
\begin{align}\label{eq:Silv-F_b}
    F_b = &\int d^3\mathbf{r}  \left\{\; f_c^B - \alpha \textrm{Tr}\left(\mathbf{AA}^\dagger\right) + \beta_1\left|\textrm{Tr}\left(\mathbf{AA}^T\right)\right|^2 \nonumber \right.\\
    &+ \beta_2\left[\textrm{Tr}\left(\mathbf{AA}^\dagger\right)\right]^2  +\beta_3\textrm{Tr}\left(\mathbf{AA}^T\mathbf{A}^*\mathbf{A}^\dagger\right)\\
    &+ \left. \beta_4\textrm{Tr}\left(\mathbf{AA}^\dagger\mathbf{AA}^\dagger\right) + \beta_5\textrm{Tr}\left(\mathbf{AA}^\dagger\mathbf{A}^* \mathbf{A}^T\right)\right\}, \nonumber
\end{align}
where we denote the condensation energy density of the isotropic bulk B phase $f_c^B$, and $F_g$ the gradient energy:
\begin{align}
    F_g = K\int d^3\mathbf{r} \left[\left(\gamma-1\right)\nabla_i A_{\mu i}^*\nabla_j A_{\mu j} + \nabla_i A_{\mu j}^* \nabla_i A_{\mu j}\right]. \label{eq:Silv-FG}
\end{align}
Near $T_c$ and at low pressure, the input parameters $\alpha$, $\beta_i$, $\gamma$, $K$, may be calculated using weak-coupling theory as:
\begin{align}
    \alpha(T, P) &=N(0)\left(1-T/T_c(P)\right)/3, \label{eq:WC-alpha} \\
    \beta_0(P) &= \frac{7\zeta(3)}{240 \pi^2} \frac{N(0)}{\left(k_B T_c(P)\right)^2},\\
    \beta_i(P) &=  n_i \beta_0(P),\quad \quad n_i = (-1, 2, 2, 2, -2)\\
    K(P) &= \frac{7\zeta(3)}{60}N(0)\left(\frac{\hbar v_F}{2\pi k_B T_c(P)}\right)^2, \label{eq:WC-K1}\\
    K(P)&\equiv\alpha(T, P)\,\xi_{GL}^2, \label{eq:WC-Kdef}\\
    \xi_{GL}(T, P) &= \frac{\xi_0}{\sqrt{1-T/T_c}}\left(\frac{7 \zeta(3)}{20}\right)^{1/2}, \label{eq:GL-coherence}\\
    \xi_0(P) &= \frac{\hbar v_F}{2\pi k_B T_c(P)}, \label{eq:zero-t-length}\\
    \gamma = 3, \label{eq:gamma}
\end{align}
where $N(0)$ is the single-spin density of states at the Fermi energy in the normal state, $v_F$ is the Fermi velocity, $\xi_{GL}$ is the Ginzburg-Landau coherence length, and $\xi_0$ is the zero-temperature coherence length. These coherence lengths serve as the characteristic distance increment for the system at a given temperature and pressure. To determine the equilibrium states of the system we find the Euler-Lagrange equations of the functional varying about a spatially-dependent order parameter $A_{\mu i}(z)$ of the form:
\begin{align}
A_{\mu i}(z) = 
    \begin{bmatrix}
    \Delta_{xx}(z) & 0 & 0 \\
    \addlinespace[1ex]
    0 & \Delta_{yy}(z) & 0 \\
    \addlinespace[1ex] 
    0 & 0 & \Delta_{zz}(z)
    \end{bmatrix} \label{eq:free-B-phase},
\end{align}
where, for consistency with confinement nomenclature, we have chosen our order parameter to be varying in $\hat{z}$. We consider the order parameter in $\hat{x}$ and $\hat{y}$ to be spatially homogeneous and infinite $(\nabla_{x,y} A_{\mu i}\equiv 0)$. By applying the boundary conditions in Table \ref{table-order-parameters}, we can solve for any domain wall configuration, provided the spatial variation is along only one axis. Equation (\ref{eq:Silv-FG}) shows that the solutions for $A_{\mu z}$ with $\nabla_z \ne 0$ are mathematically equivalent to solutions for $A_{\mu x}$ with $\nabla_x \ne 0$. Therefore, it is possible to rotate the solution axes by switching order parameter component.

To facilitate the numerical work, we employ a scaled order parameter $a_{\mu i} = A_{\mu i} / \Delta_B$ proposed in Ref.~\cite{li1988}, where $\Delta_B$ is the equilibrium $B$-phase order parameter in the bulk superfluid. From weak-coupling theory, we compute the bulk order parameter and scaling coefficients:
\begin{align}
\Delta_B &= \left[\frac{\alpha}{2(3\beta_{12} + \beta_{345})}\right]^{1/2}, \label{eq:bulk-B-OP} \\
\partial_i &\equiv \xi_{GL} \,\nabla_i,  \label{eq:def-partial}\\
\zeta_i &\equiv \frac{\beta_i}{3\beta_{12} + \beta_{345}}, \label{eq:def-zeta}
\end{align}
where we have used the convention $\beta_{ij\cdots}\equiv \beta_i + \beta_j + \ldots$ Defining a dimensionless free energy density $\mathfrak{f} \equiv f / (\alpha \, \Delta_B^2)$, we obtain expressions for the bulk and gradient contributions:
\begin{align}
    \mathfrak{f}_B = &  \tfrac{3}{2} - \textrm{Tr}\left(\mathbf{aa}^\dagger\right) + \tfrac{1}{2} \zeta_1\left|\textrm{Tr}\left(\mathbf{aa}^T\right)\right|^2 \nonumber 
    + \tfrac{1}{2}\zeta_2\left[\textrm{Tr}\left(\mathbf{aa}^\dagger\right)\right]^2 \\ &+\tfrac{1}{2}\zeta_3\textrm{Tr}\left(\mathbf{aa}^T\mathbf{a}^*\mathbf{a}^\dagger\right)\label{eq:fb-scaled} + \tfrac{1}{2}\zeta_4\textrm{Tr}\left(\mathbf{aa}^\dagger\mathbf{aa}^\dagger\right)\\ 
    &+\tfrac{1}{2}\zeta_5\textrm{Tr}\left(\mathbf{aa}^\dagger\mathbf{a}^* \mathbf{a}^T\right), \nonumber\\
    \mathfrak{f}_G = &(\gamma - 1)\partial_j a_{\mu j} \partial_i a_{\mu i}^* + \partial_j a_{\mu i} \partial_j a_{\mu i}^*, \label{eq:fg-scaled}
\end{align}
and a total dimensionless free energy $\mathfrak{F} = \int \left(\mathfrak{f}_B + \mathfrak{f}_G\right)$.

\begin{figure}[t]
\includegraphics[width=\columnwidth]{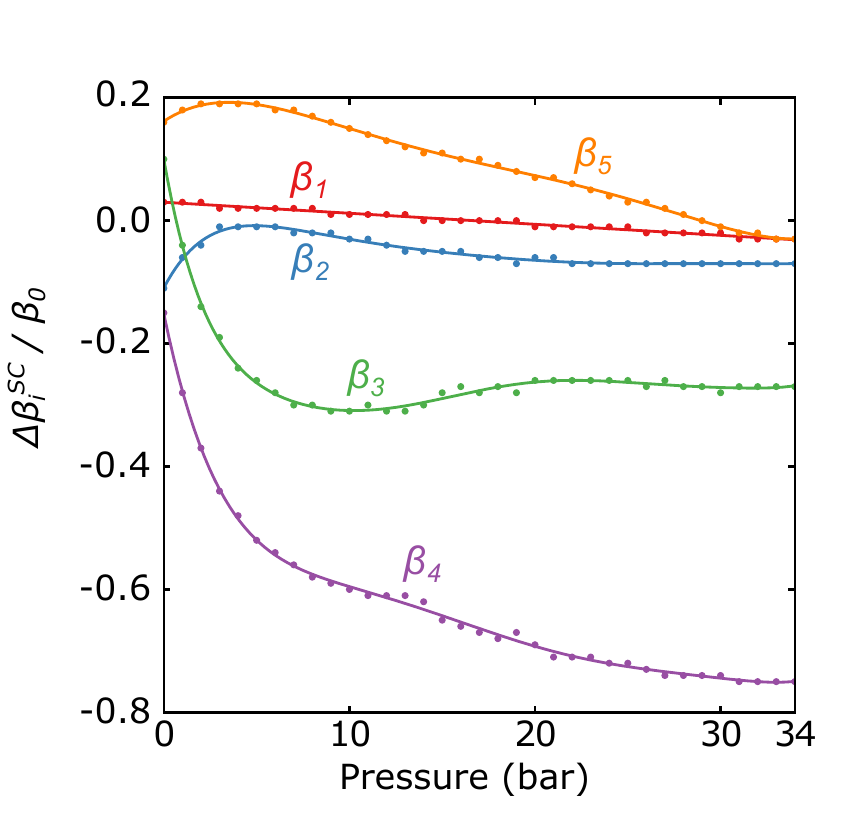}
\caption{$\Delta \beta_i^{SC}$ values from \cite{choi2007} \cite{choi2013}, with polynomial fits used in this work.}
\label{fig:choi-betas}
\end{figure}

\subsection{Strong-coupling corrections}

The weak-coupling values of the parameters $\alpha$, $\beta_i$, $\gamma$, $K$, are only expected to give valid results in the low-pressure regime, as the pairing interaction itself is a functional of the order parameter \cite{wolfle1974}. To expand our results to the pressure range $P \in [0, 34]$ bar, we implement experimentally-obtained corrections. Employing the Doniach-Engelsberg exchange-enhancement model \cite{doniach1966}, we do not expect there to be corrections to the second-order terms $K$, $\alpha$, $\gamma$ and we continue to use values calculated using weak-coupling theory. These corrections give strong-coupling parameters:
\begin{align}
    \beta_i(T, P) = \beta_i^{WC}\left(P, T_c(P)\right) + \frac{T}{T_c}\Delta\beta_i^{SC}(P), \label{eq:SC-corrections}
\end{align}
which allow us to extend our calculations over the experimentally-measured pressure domain. To increase the resolution of our calculations, we apply polynomial fits. Experimental data from Refs.~\cite{choi2007, choi2013}, along with polynomial fits, may be seen in Fig.~\ref{fig:choi-betas}.

GL theory is based on an expansion of the free energy about a small order parameter and is only expected to be valid for temperatures near $T_c$. To allow the order parameter to saturate at lower temperatures, we follow the example of Ref.~\cite{wiman2016} and replace the Ginzburg-Landau coherence length $\xi_{GL}$ in (\ref{eq:WC-Kdef}) by the characteristic length scale from weak-coupling BCS theory:
\begin{align}
    \xi_\Delta(T) &= \frac{\hbar v_F}{\sqrt{10}\,\Delta_B^{\textrm{BCS}}(T)}, \label{eq:xi-delta}
\end{align}
which is consistent with $\xi_{GL}$ in the regime $T\approx T_c$. Using the approximation \cite{carless1983}:
\begin{align}
    \frac{\Delta(T)}{\Delta(0)} \approx 
        \begin{cases}
            \sqrt{1 -T/T_c}\,(0.9663 + 0.7733 T/T_c),& t\ge 0.3 \label{eq:Delta-T}\\ 
            1,& t<0.3 
        \end{cases},
\end{align}
for the reduced temperature $t=T/T_c$, and $\Delta(0) \approx 1.76 \,k_B T$ in $^3$He-$B$ \cite{vollhardt2013}, we can set up a relationship between these lengths which will allow for an accurate determination of the spatial variation of the order parameter for temperatures $T<T_c$.

\subsection{Interfacial tension}
To determine the relative cost of the different structures examined, we exploit the fact that our system is translationally invariant in two dimensions and define the tension, following Ref.~\cite{silveri2014}:
\begin{align}
    \sigma = \frac{\mathfrak{F}}{A}, \label{eq:sigma}
\end{align}
where $A$ is the area of the interface. Thus two of the dimensions in the spatial integral cancel exactly and we need only integrate equations (\ref{eq:fb-scaled}) and (\ref{eq:fg-scaled}) in one dimension. This integration will introduce a factor of the characteristic length scale and therefore suggests scaling by a factor of $f_c^B \xi_\Delta$, which in our scaled energy functional (\ref{eq:fb-scaled}) is exactly -1.5, in agreement with Ref.~\cite{silveri2014}. Numerical integration of the condensation energy in this manner returns -1.5 to better than one part in $\num{3d15}$. We use the theoretical value in these calculations to avoid singularities when determining the gradient contribution alone. The integrations were performed numerically using the converged solutions for the order parameter and a standard Python library quadrature integration routine. Weak-coupling values for the interface tension are given in Table \ref{table-order-parameters}.

\begin{figure}[htbp]
\includegraphics[width=\columnwidth]{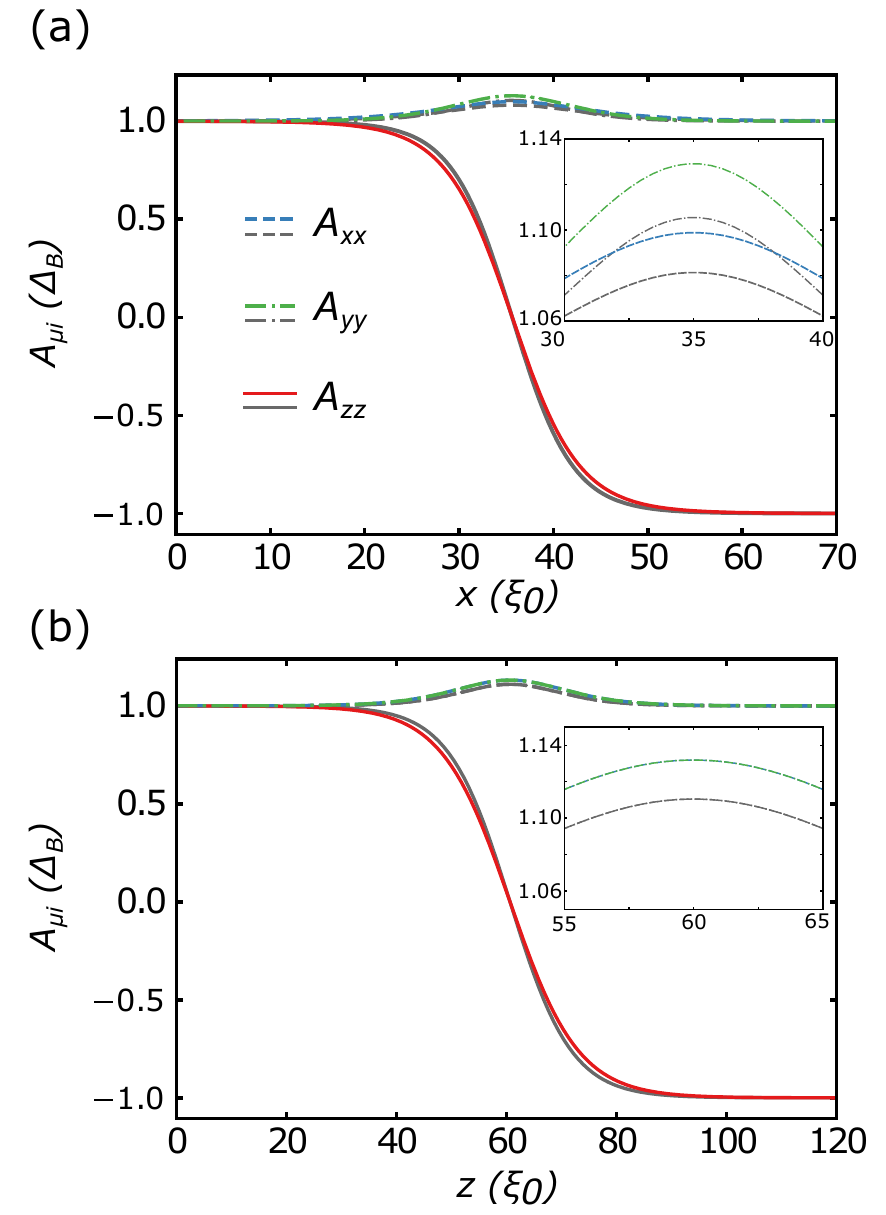}
\caption{Converged strong-coupling solution to the three nonzero components of the order parameter for domain walls at $T=0.85\,T_c(15\,\textrm{bar})$, $P=$ 15 bar. Interfaces presented are (a) type-$x$ ($BB10$), and (b) type-$z$ ($BB\overline{1}2$). Weak-coupling solutions are underlaid in grey. Insets show an enlarged area of $A_{xx}$, $A_{yy}$ near the domain wall. Note the separation between components in (a) due to the higher gradient energy contribution of $A_{xx}$ causing suppression in the hard axis. This separation is not present in the type-$z$ interface as both components have equally weighted gradient energies.}
\label{fig:tall-trans-spatial}
\end{figure}
\begin{figure*}[ht]
\includegraphics[width=\textwidth]{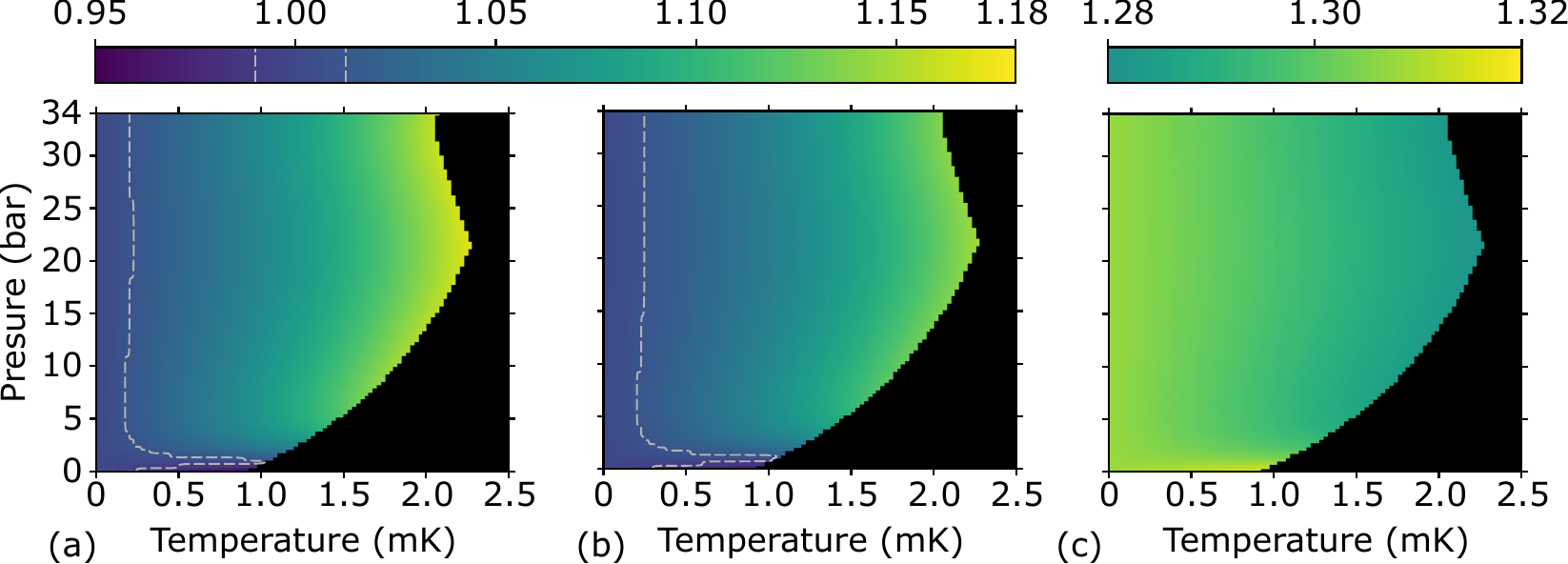}
\caption{Domain wall widths, $l\,(\xi_\Delta)$, for the $BB10$ (type-$x$) and $BB\overline{1}2$ (type-$z$) interfaces, showing that strong-coupling corrections increase the domain wall width over most of the phase diagram. Grey dashes in (a), (b), and their colorbar mark the boundary of the region in the phase diagram where the weak-coupling approximation applies. Temperatures below the dashed line agree with the weak-coupling result within $\pm 1\%$. (a) Ratio of strong-coupling to weak-coupling domain size for the $BB10$ interface. (b) Ratio of strong-coupling to weak-coupling domain size for the $BB\overline{1}2$ interface. We see that the thickness of this domain wall is not increased with strong coupling as much as for the $BB10$ interface. Though not plotted here, this trend is consistent across all interfaces examined. (c) Ratio of strong-coupling values $BB\overline{1}2 / BB10$. The type-$z$ interface is universally wider than the type-$x$ interface. This is the case for all interfaces, with $BB10$ universally the narrowest domain wall.} 
\label{fig:ratio-domains-b12/10}
\end{figure*}
\subsection{Stability and decay}
Multiple self-consistent solutions to the minimization of the Ginzburg-Landau free-energy functional  (equations (\ref{eq:fb-scaled}), (\ref{eq:fg-scaled})), are possible. We employ a relaxation method, wherein an initial solution guess is iterated upon until the residuals from the Euler-Lagrange equations converge within a specified tolerance. Each iteration approaches the final solution more closely, and this process may be seen to be analogous to iteration towards lower-energy states of the system \cite{thuneberg1987}. It follows that the obtained solution is highly dependent on the shape of the initial mesh and it is possible that the solutions we find do not represent global energy minima. Alternative solutions to the boundary value problem may represent lower-energy, physically realizable systems \cite{silveri2014}. To access these lower-energy states, we follow the example of Ref.~\cite{silveri2014} and apply a series of perturbations to the converged solution. These perturbations were added to a converged solution and used as a new initial guess for the solver. The primary perturbations we employed were of the form:
\begin{align}\label{eq:perturb_invcosh}
    \delta A_{\mu i}(z) = \sum_{k=0}^N \frac{C^{(k)}_{\mu i}}{\cosh\left[\left(z - z_k\right)/s\right]}
\end{align}
with parameters $C$, $z_k$, $N$, and $s$. The most important parameter for our analysis is the perturbation amplitude $C$:
\begin{align}
    C = \max_{\mu, i} \delta A_{\mu i}(z).
\end{align}
The other parameters; $z_k$ the spatial shift of the perturbation, $N$ the number of simultaneous perturbations, and $s$ the width parameter of the perturbation; were changed in a situation-dependent manner, with a variety of combinations applied to each interface. Perturbations were applied to (up to) all nine real components of the converged order parameter to provoke varying decays. To test the dependence of the converged solution on the type of perturbation applied, alternative trigonometric perturbations were tested. When applied with comparable amplitude, spatial shift, and widths, these were found to return solutions consistent with the inverse hyperbolic perturbations in equation (\ref{eq:perturb_invcosh}).

Our stability analysis comprised of finding the lowest-amplitude perturbation which could cause an alternative solution to converge in weak coupling, then establishing whether this alternative solution would still be formed in the strong-coupling regime. For all decay schemes explored, we found no difference in this critical perturbation amplitude between weak and strong coupling, with the same alternative solutions being formed in each case. We are able to access many of the same decay schemes reported in Ref.~\cite{silveri2014}, though we note that the accessible solutions as well as the exact perturbation required to obtain differing solutions necessarily vary by solver algorithm. Our analysis gives no evidence contradicting the stability conclusions of Ref.~\cite{silveri2014}, namely that $BB10$ is stable against real perturbation, while all other interfaces examined decay more easily to alternative domain wall/texture solutions. Notably, we confirm that $BB\overline{1}2$ remains susceptible to decay into either a double interface or a texture, dependent on the form of the perturbation. This lack of stability in the bulk, however, does not preclude the type-$z$ domain wall's characteristic suppression of $A_{zz}$ as it transits from 1 to -1. As such it remains a good choice for modeling the surface pairbreaking at a confinement boundary, as discussed later in Section \ref{sec:PDW geometry}. These calculations allow us to conclude that domain wall stability is not affected by strong-coupling corrections.

\section{Spatial variations of the order parameter}\label{sec:spatial-var}
We have performed simulations of the order parameter's variation across an arbitrary domain of $D = 140 \, \xi_{\Delta}$, following Ref.~\cite{silveri2014}, with Dirichlet boundary conditions and the final form of the order parameter detailed in Table \ref{table-order-parameters}. Simulations were performed using both weak and strong-coupling parameters in the parameter space $T \in [0, \,T_c(P)]$, $P \in [0, \,34]$ bar. Solutions were obtained using standard Python libraries for boundary-value problems, using either unity or Fermi functions as the initial mesh, depending on the boundary conditions. We find that our weak-coupling results are consistent with those reported in Ref.~\cite{silveri2014}, confirming the validity of our computational method.

Figure \ref{fig:tall-trans-spatial} shows the change in order parameter for the $BB10$ and $BB\overline{1}2$ domain walls at 15 bar and $T=0.85\,T_c$. We see that in both types, the domain wall with strong-coupling corrections is thicker, and that the components of the order parameter with unchanged sign grow --- as shown in the insets. The increase in domain-wall thickness is observed over the majority of the phase diagram as shown in Fig.~\ref{fig:ratio-domains-b12/10}. We also observe an increase in domain-wall width with increasing temperature, with a larger change in the $BB10$ interface than in the $BB\overline{1}2$ interface. This trend is seen at all pressures except a small region near saturated vapor pressure, which shows the opposite trend, Fig.~\ref{fig:tall-trans-spatial}. Note that the weak-coupling solutions are necessarily uniform throughout the parameter space and are reported in Table \ref{table-order-parameters}. 

\begin{figure*}[ht]
\includegraphics[width=\textwidth]{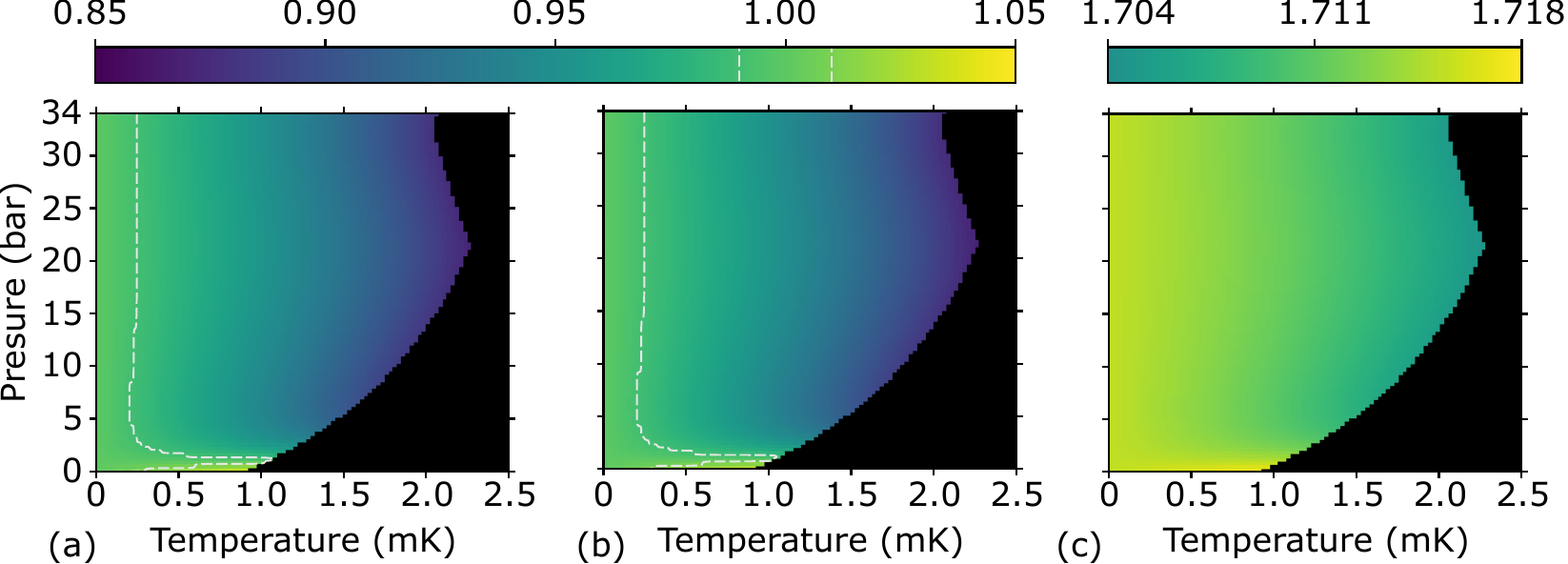}
\caption{Interfacial tension, $\sigma\,(f_c^B \xi_\Delta)$, for the $BB10$ (type-$x$) and  $BB\overline{1}2$ (type-$z$) interfaces, showing that the strong-coupling corrections lower the interface tension across most of the phase diagram. Only the two simplest interfaces are plotted here, however the trend observed in (a) and (b) is consistent for all domain walls studied. Grey dashes in (a), (b), and their colorbar mark the boundary of the region in the phase diagram where the weak-coupling approximation applies. Temperatures below the dashed line agree with the weak-coupling result within $\pm 1\%$. (a) Ratio of strong-coupling to weak-coupling tension for the $BB10$ interface. (b) Ratio of strong-coupling to weak-coupling tension for the $BB\overline{1}2$ interface. (c) Ratio of strong-coupling values $\sigma_z / \sigma_x$, showing that the $BB10$ interface is universally lower-energy than the $BB\overline{1}2$ interface. This is consistent for all domain wall configurations with $BB10$ always seen to be lower-energy.} 
\label{fig:ratio-tensions-b12/10}
\end{figure*}

In Fig.~\ref{fig:ratio-domains-b12/10}(c) we compare the type-$x$ and type-$z$ solutions. We observe that the width of the $BB\overline{1}2$ wall is always greater than that of the $BB10$ wall, with the solutions approaching one another at increased temperature and pressure. We also note that the ratio between the $BB\overline{1}2$ wall and the $BB10$ wall is always greater than that between the weak and strong-coupling solutions. While this result may follow naturally in weak coupling from (\ref{eq:Silv-FG}) and (\ref{eq:gamma}), its persistence in the presence of strong-coupling corrections indicates that, spatially, the effect of changing domain wall types is more significant than changing from weak to strong-coupling $\beta$ parameters. 
In light of these results, we do not expect the characteristic length scales of physically-implementable systems to be significantly altered by strong-coupling corrections.
 
\section{Interfacial Tension}\label{sec:tension}

We gauge the relative likelihood of domain wall formation by computing the interface tension $\sigma$ across the same phase diagram that we used for our analysis of the spatial deformation. This value serves as an indicator of the energy cost of the hard domain wall as compared to the bulk condensation energy. For these calculations, the bulk $B$ phase has $\sigma = 0$. As in section \ref{sec:spatial-var}, we use Ref.~\cite{silveri2014} as a comparison to confirm the validity of our calculations \footnote{We note that in Ref.~\cite{silveri2014} the reported values for $\sigma$ are for their perturbed minimal solutions. We are interested in the tensions for the single domain walls and report values for our initial converged solutions.}. As observed for the spatial deformation, weak-coupling results are necessarily $P$, $T$-independent and are reported in Table \ref{table-order-parameters}.

Selected results for the interface tension are presented in Fig.~\ref{fig:ratio-tensions-b12/10}. Taking the ratio of the tensions $\sigma_{SC} / \sigma_{WC}$, we see that including the strong-coupling corrections reduces the interface tension across most of the phase diagram with the exception of the anomalous low-pressure region also observed in Fig.~\ref{fig:ratio-domains-b12/10}. The relative energy cost of the domain wall is reduced as $T\rightarrow T_c$ with the lowest tensions calculated in the region where the domain wall is thickest. Even for the most energetically costly $BB\overline{12}$ interfaces, which have $\sigma_{WC}$ over an order of magnitude greater than $BB10$, the ratio $\sigma_{SC}/\sigma_{WC}$ remains consistently $\in [0.85, 1.15]$ with a near-identical distribution, providing further evidence that the shift between domain wall types is much more significant than the shift between weak and strong coupling. When we examine the tension ratio between domain wall types as in Fig.~\ref{fig:ratio-tensions-b12/10}(c), we see that temperature and pressure have negligible effects on this value. That is, strong-coupling corrections affect all domain walls similarly.

Two areas of the phase diagram are of particular interest. Close examination of the region near saturated vapor pressure reveals that the trend of increasing interfacial tension with decreasing pressure continues past unity (the weak-coupling approximation) and gives an \textit{increased} domain wall tension for pressures below 0.7 bar. This region of increased energy of the $BB10$ domain wall, in particular, suggests that searches for the PDW state should not be focused on the low-pressure region of the phase diagram. Additionally, examining Fig.~\ref{fig:ratio-tensions-b12/10}, we note that the high-temperature region where the interfacial tension is lowest in panels (a), (b), coincides with the region in panel (c) where the tension ratio is lowest. This suggests to us that the strong-coupling corrections contribute differently between domain wall types and spurred further investigation into the energetics of domain walls.

To investigate these differences, we recall definitions of $\mathfrak{F}$ and $\sigma$ from equations (\ref{eq:fb-scaled}), (\ref{eq:fg-scaled}) and (\ref{eq:sigma}) we perform the separation:
\begin{align}
    \sigma &= \sigma_B + \sigma_G,
\end{align}
which allows us to examine the relative contributions of the gradient and bulk components of the interfacial tension. Taking the ratio of $\sigma_G / \sigma$, we note that the gradient contribution to the interfacial tension of the domain wall is relatively constant across the phase diagram, with variations on the order of the variation in Fig.~\ref{fig:ratio-tensions-b12/10} (c). Additionally, the changes from weak to strong-coupling gradient ratios are negligible, on the order of 0.01\%. We also confirm, as detailed in Table \ref{table-order-parameters}, the gradient contributions are almost exactly one-half of the total interface tension regardless of domain wall type, as originally calculated in \cite{salomaa1988}. Since the gradient energy is not affected by strong-coupling corrections in our model, changes in the bulk free-energy calculation (\ref{eq:Silv-F_b}) due to strong-coupling corrected $\beta$ parameters must be perfectly matched by concomitant changes to the spatial solution in order to preserve this energy balance. The equality between gradient and bulk contributions can be explained by energy-conservation arguments \cite{salomaa1988}, and thus is not expected to be altered in strong coupling. The gradient contribution may be affected by any confinement effects for the type-$z$ domain walls, and it would be interesting to further investigate the implications of this energy breakdown for the formation of PDW states in confined geometry. 

The general decrease in interfacial tension across the phase diagram, combined with our results showing that strong-coupling corrections contribute less to domain wall energetics than the domain wall's type, strongly suggest that the mechanisms for the formation of hard domain walls, and therefore PDW states, will not be suppressed in strong-coupling.

\section{Domain Walls in a PDW}\label{sec:PDW geometry}
\begin{figure}
    \centering
    \includegraphics[width=\columnwidth]{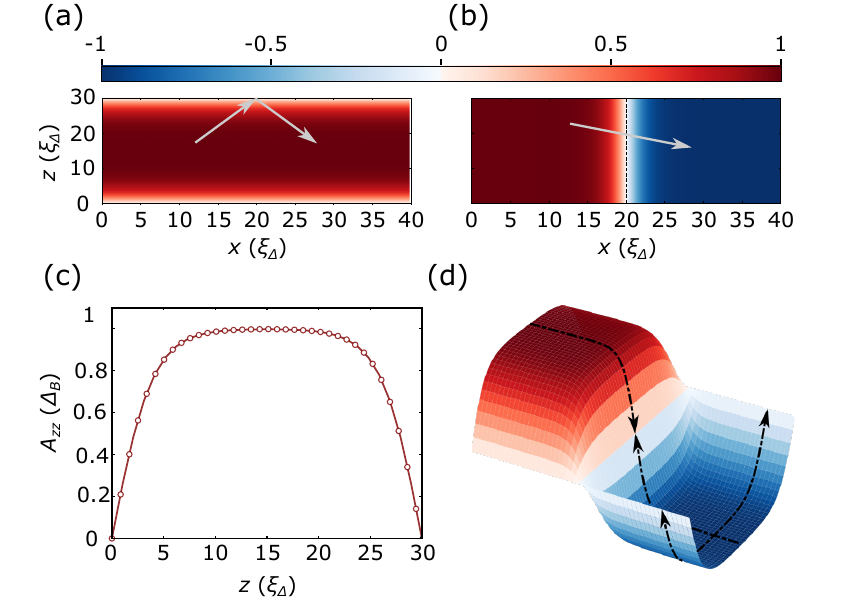}
    \caption{Schematic representation of hard domain walls in confined geometry. (a)  1D type-$z$ solution projected onto 2D geometry confined in $\hat{z}$. The maximal pairbreaking trajectory is indicated by the grey arrow, with the order parameter component $A_{zz}$ suppressed at the domain walls. (b) 1D type-$x$ solution projected onto 2D geometry confined in $\hat{z}$. We perform this projection by rotating axes as described in Section \ref{sec:PDW geometry}, presenting the results as varying in $\hat{x}$ to facilitate the visualization of pairbreaking in a PDW. The pairbreaking trajectory in (a) is compensated for by the sign change in $A_{zz}$ across the domain wall. This domain wall creates an additional pairbreaking trajectory across the wall, indicated by the grey arrow. (c) Type-$z$ solutions (open circles) overlaid with a direct solution for confined geometry and specular boundary conditions (solid line), as calculated in Ref.~\cite{li1988}, confirming the validity of mapping a type-$z$ solution onto this geometry. (d) 3D representation of the superposition of these domain walls into a PDW-like state. Each dashed arrowhead represents one-half of a domain wall solution, with the head pointing towards the node in $A_{zz}$.}
    \label{fig:mapping-pairbreaking}
\end{figure}
The original prediction of the stripe phase comprised of broken gauge and translational symmetry, with hard domain walls separating a periodic arrangement of degenerate domains of $^3$He-$B$ \cite{vorontsov2005, vorontsov2007}. Since this translational symmetry-breaking occurs in the plane of the confined slab, it is necessarily a two-dimensional system. However, the proposed pair-breaking interactions are well-described by domain walls in 1D, the subject of this work.  

The pair-breaking suppression of $A_{\mu z}$ at a confinement boundary may be modeled using the $BB\overline{1}2$ domain wall.  The order parameter $A_{zz}$ goes to zero at the node of the solution, and rearrangement of this curve as shown in Fig. \ref{fig:mapping-pairbreaking} (a) shows this surface scattering effect. This is not just a qualitative similarity, as the rearranged type-$z$ domain wall conforms exactly to solutions of the planar-distorted $B$ phase in specular boundary conditions,   computed using the method of Ref.~\cite{li1988}. We show these solutions superimposed in Fig.~\ref{fig:mapping-pairbreaking} (c). It is proposed that alternating $BB10$ domain walls across the direction of confinement would reduce the pair-breaking cost in regions near the domain wall by forcing the sign change across the entire domain wall \cite{vorontsov2007, wiman2015, wiman2016, vorontsov2018, yapa2021}. This type-$x$ pair-breaking trajectory is contrasted with the surface interaction in Fig.~\ref{fig:mapping-pairbreaking} (b). The PDW state may be considered to be a combination of these domain walls. In Fig.~\ref{fig:mapping-pairbreaking} (d), we take the product of the type-$x$ and type-$z$ solutions to give an approximate representation of $A_{zz}$ in such a two-dimensional configuration.

\section{Conclusion}
We have applied experimental strong-coupling corrections to Ginzburg-Landau free-energy calculations of hard domain walls in superfluid $^3$He-$B$. The self-consistent solutions to the order parameter were found to have increased domain-wall thickness and variation in amplitude when compared to weak-coupling solutions. Interface tensions were computed for each domain wall configuration and were found in all cases to lower the interfacial tension across the majority of the phase diagram. We note that variation across the phase diagram for a single domain wall is near-universally smaller than variation caused by changing interface configuration and thus that domain wall type is the greatest determiner of interface energetics, exceeding the effects of temperature, pressure, and thus strong-coupling corrections. We examine the relative contributions of gradient and bulk interfacial tension and find that the gradient contribution is constantly one-half of the total tension in all domain wall types. From these results, we conclude that domain wall formation, and the PDW state, should not be adversely affected by strong coupling.

\begin{figure}[ht]
    \centering
    \includegraphics[width=0.8\columnwidth]{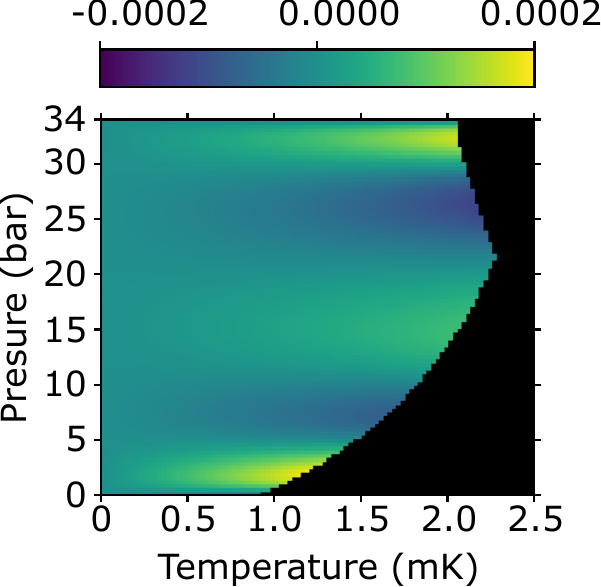}
    \caption{Differences between calculations of $\sigma_z/\sigma_x$ in strong coupling, performed using Wiman $\Delta\beta_i$ parameters \cite{wiman2018, regan2020} versus Choi $\Delta \beta_i$ parameters \cite{choi2007}, as described in the appendix. Here, the Choi calculations have been subtracted from the Wiman calculations. This phase diagram is representative of these differences for all the calculations performed in our manuscript, with only the magnitude of the differences changing. The extrema of these differences for each calculation are presented in Table \ref{table-appx-summ}.}
    \label{fig:appendix-repr}
\end{figure}

\section{Acknowledgments}
%In next version of this manuscript, correct acknowledgements if necessary.
This work was supported by the University of Alberta; the Natural Sciences and Engineering Research Council, Canada (Grants No. RGPIN-04523-16 and No. CREATE-495446-17); and the Alberta Quantum Major Innovation Fund. J.M. was supported by NSERC Discovery Grants Nos. RGPIN-2020-06999 and RGPAS-2020-00064; the CRC Program; CIFAR; a Tri-Agency NFRF Grant (Exploration Stream); and the PIMS CRG program.

\section*{Appendix: Alternative correction parameters}\label{sec:appendix}
During the review process, we were made aware of new calculations of $\Delta \beta_i$ \cite{sauls2021}, performed by J.J.~Wiman in his PhD thesis \cite{wiman2018} and which are published in Ref.~\cite{regan2020}. These calculations are based on new microscopic theory and represent an updated, as well as potentially more accurate, strong-coupling parameter set. To resolve any ambiguity, we have since performed our calculations using these new parameters.

We find no alterations to our conclusions, with the new results for both domain wall thickness and interface tension calculations in agreement with our prior results within at least 0.1\%, and for some calculations we find significantly smaller changes. The plots generated by these new parameters are visually indistinguishable from our Figs.~\ref{fig:ratio-domains-b12/10}, \ref{fig:ratio-tensions-b12/10}, and as such we have not included them here. Instead, we have computed the differences between results using both sets of $\Delta\beta_i$ parameters and present them in this appendix. The principal differences are located in areas near $T_c$, as would be expected due to the form of the corrections in Eqn.~(\ref{eq:SC-corrections}). We note that these differences are minute and do not affect the trends described in Sections \ref{sec:spatial-var} and \ref{sec:tension}. Most significantly, the differences computed in the plots comparing domain wall types (Fig.~\ref{fig:appendix-repr}, Table \ref{table-appx-summ}) are an order of magnitude smaller than those for the single domain walls, as seen in Table \ref{table-appx-summ}. This reduction in extrema further reinforces our conclusion that these revised parameters do not affect the primary conclusion of this manuscript --- namely that differences between domain walls are far more significant than those imparted by strong-coupling corrections. We further note that the zone of decreased thickness/increased tension present near saturated vapor pressure is preserved with these new parameters, despite the markedly different low-pressure behavior of the Wiman $\Delta \beta_i$ parameters. Thus, by having repeated these calculations with the Wiman parameters \cite{regan2020} we feel our results are reinforced, and continue to expect domain wall formation to be unaffected in the strong-coupling regime.
\begin{table}[hb]
\begin{ruledtabular}
\begin{tabular}{ccccc}
\addlinespace[1ex]
Calculation & Equiv. Fig. & Difference Extrema\\
\addlinespace[1ex]
\colrule
\addlinespace[2ex]
$l_{SC}/l_{WC}$ $(BB10)$ & \ref{fig:ratio-domains-b12/10}(a) & $\num{-3.01d-3},\,\num{2.00d-3}$\\
\addlinespace[1.5ex]
$l_{SC}/l_{WC}$ $(BB\overline{1}2)$ & \ref{fig:ratio-domains-b12/10}(b) & $\num{-2.74d-3},\,\num{1.79d-3}$ \\
\addlinespace[1.5ex]
\multicolumn{1}{l}{$\;l_z/l_x$ (SC)} & \ref{fig:ratio-domains-b12/10}(c) & $\num{-3.45d-4},\,\num{6.43d-4}$\\
\addlinespace[1.5ex]
$\sigma_{SC}/\sigma{WC}$ $(BB10)$ & \ref{fig:ratio-tensions-b12/10}(a) & $\num{-1.35d-3},\,\num{2.37d-3}$\\
\addlinespace[1.5ex]
$\sigma_{SC}/\sigma{WC}$ $(BB\overline{1}2)$ & \ref{fig:ratio-tensions-b12/10}(b) & $\num{-1.40d-3},\,\num{2.47d-3}$\\
\addlinespace[1.5ex]
\multicolumn{1}{l}{$\sigma_{z}/\sigma{x}$ $(SC)$} & \ref{fig:ratio-tensions-b12/10}(c) & $\num{-1.18d-4},\,\num{1.88d-4}$\\
\end{tabular}
\end{ruledtabular}
\caption{\label{table-appx-summ}Summary of differences between use of $\Delta\beta_i$ parameters from Ref.~\cite{wiman2018,regan2020} versus from Ref.~\cite{choi2007}. The first column is the mathematical operation performed, with the second column denoting the figure in which this calculation is presented in the main body of this manuscript. The third column gives the extrema of the differences between each set of calculations, with the Choi calculations having been subtracted from the Wiman calculations.}
\end{table}

\balance

\nocite{*}

\bibliography{references.bib}% Produces the bibliography via BibTeX.

\end{document}